\documentclass[10pt,twocolumn]{article}
\usepackage{amssymb,amsmath}
\usepackage{amsfonts}
\usepackage{mathtools}
\usepackage[super]{nth}
\usepackage[letterpaper, margin=.75in]{geometry}
\usepackage{lipsum}
\usepackage{authblk}

\usepackage{natbib}
\usepackage{graphicx}
\begin{document}

\title{\vspace{-2pc}Modeling the Ionosphere with GPS and Rotation
Measure Observations\vspace{-1ex}}

 \author[1]{J. B. Malins}
 \author[2]{S. M. White}
 \author[1]{G. B. Taylor}
 \author[3]{K. Stovall}
 \author[1]{J Dowell}
 

\affil[1]{Department of Physics and Astronomy,
University of New Mexico, Albuquerque, Arizona, USA.}
\affil[2]{Air Force Research Laboratory, United States Air Force,
Kirtland Air Force Base, New Mexico, USA.}
\affil[3]{Very Large Array, National Radio Astronomy Observatory, Socorro, New Mexico, USA.}

\twocolumn[
    \begin{@twocolumnfalse}
    \maketitle
    \begin{abstract}
        The ionosphere contributes time-varying Faraday Rotation (FR) to radio signals passing through it. Correction for the effect of the ionosphere is important for deriving magnetic field information from FR observations of polarized cosmic radio sources, as well as providing valuable diagnostics of the structure of the ionosphere. In this paper, we evaluate the accuracy of models commonly used to correct for its effects using new observations of pulsars at low frequencies, which provide total rotation measures (RM) at better precision than previously available. We evaluate models of the ionosphere derived from modern digital ionosondes that provide electron density information as a function of height, as well as GPS-derived Total Electron Content (TEC) measurements. We combine these density models with reference global magnetic field models to derive ionospheric RM contributions. We find that the models disagree substantially with each other and seek corrections that may explain the differences in RM prediction. Additionally we compare these models to global TEC models and find that local high-cadence TEC measurements are superior to global models for ionospheric RM correction.
    \end{abstract}
    \end{@twocolumnfalse}
    \vskip 1.0cm
]

\section{Introduction}

The ionosphere is the layer of the Earth's upper atmosphere in which particles are charged due to ionization by short wavelength (ultraviolet, EUV, XUV and X-ray) radiation from the Sun. As a magnetoactive plasma, the ionosphere has a significant impact on incoming radio signal propagation from space, whether from nearby satellites or distant cosmic sources. The effects of the ionosphere are typically much more pronounced at longer wavelengths, and hence accurate modelling of ionospheric effects is important for the new class of sensitive, long-wavelength radio telescopes such as the Long Wavelength Array (LWA) \citep{LWARef}, the Murchison Widefield Array (MWA) \citep{MWARef} and the Low Frequency Array (LoFAR) \citep{LOFARRef}. These telescopes are affected both by electron density fluctuations in the ionosphere and by Faraday rotation, which can be used to probe the ionosphere by measuring effects on observations of distant radio sources. 

Faraday Rotation (FR) is the rotation of the plane of linear polarization of an electromagnetic wave as it passes through a magnetized plasma. The plane of polarization rotates because the wave is the sum of components in the two natural elliptically-polarized modes of the plasma. Degeneracy between the two natural modes is broken in a magnetized plasma and results in slightly different phase velocities. This difference produces a frequency-dependent rotation of the orientation of linear polarization. The amount of rotation, $\phi$, of a given linearly polarized wave at wavelength $\lambda$ is given by \citep[e.g.,][]{Spitzer}

\begin{equation}
\phi\, =\, {\textrm{RM}} \ \lambda^2 
\label{eq:RM}
\end{equation}

\noindent where the quantity

\begin{equation}
\textrm{RM} = \frac{e^2}{8 \pi^2 \epsilon_0 m_e^2 c^3}\, \int_0^d n_{e}\left( s \right) \, B_{\parallel} \left( s\right), ds
\label{eq:FR}
\end{equation}

\noindent is referred to as the rotation measure (RM), or frequency-independent measure of FR. Both $n_e$, the density of electrons, and $B_\parallel$, the line of sight component of magnetic field, are functions of distance $s$ along the line of sight. The wavelength-dependence of (\ref{eq:FR}) provides a straightforward means to determine RM: one measures the orientation of the plane of linear polarization at each frequency across a sufficiently broad bandwidth and fits a $\lambda^2$-dependence \citep{Carilli}. Typical values for ionospheric RM are of order radians m$^{-2}$. The integral is carried out over the entire line-of-sight path to the radio source. For a cosmic source, this will include separate contributions from the interstellar medium (ISM), the heliosphere (the solar wind and structures within it) and the Earth's ionosphere. \citet{ObL12} discuss the relative sizes of these contributions. ISM contributions to FR generally vary slowly, and thus can readily be separated from variations due to structures in both the solar wind, such as CMEs, and the ionosphere, that can vary on timescales of tens of minutes. We can separate the heliospheric component by ensuring the pulsar is measured at a large elongations, or angle from the sun. 

In this paper we compare Long Wavelength Array Station 1 (LWA1) measurements of the FR of pulsars to model predicted FR created through ionospheric measurements combined with reference geomagnetic field information. LWA1 measurements are taken of polarized pulsars and the RM of the signal is determined directly using the RM fitting method described above. The LWA1 pulsar measurements are taken independent of any model input, allowing the models to be directly compared to actual ionospheric observations. Measurements of the pulsar are taken through the period of sunrise, and with the pulsar a large angular distance from the Sun, ensuring ionospheric changes should dominate FR variations. We compare these measurements to ionospheric models obtained from both ionosondes and GPS measurement. A new feature of this work is the use of digital ionosondes as well as GPS receivers co-located with the LWA station in order to provide ionospheric information directly comparable to the LWA1 observations. Since the strength of the geomagnetic field decreases with height, we might expect that the lower region of the ionosphere is important for determining ionospheric FR. Additionally, we use GPS measurements of the total electron column density of the ionosphere. We investigate several different models and show that the use of local TEC measurements is superior to the use of current global TEC models.

In order to compare the measurements of ionospheric FR with models, we need to calculate (\ref{eq:RM}) through the ionosphere: this requires producing profiles of electron density and line-of-sight magnetic field with height along the line of sight to the pulsar as it moves across the sky. We use the International Geomagnetic Reference Field 12 (IGRF12) \citep{Thébault2015} along with the line of sight magnetic field component of the LWA1 station to the pulsar to specify $B_\parallel$ for these models. We assume the geomagnetic field does not vary significantly with time and can be regarded as stable for our purposes. Electron density, however, varies greatly with time of day and with solar activity, with substantial density changes typically occur over the course of tens of minutes to hours. Since the electrons in the ionosphere are strongly coupled to the neutral particles at the same height, they generally exhibit smooth vertical gradients and fall into bands that are largely horizontal with characteristic horizontal gradients of order the thickness of the ionosphere or greater. The 'F' layer is the largest and densest static layer in the ionosphere, existing from 150 km to above 500 km and transitioning at its top to the plasmasphere. The F layer fluctuates in density and peak height greatly throughout the day, with a peak during the afternoon that can be as low as 200 km, while at night it can peak as high as 400 km. The F layer typically has densities of order $10^4$ electrons cm$^{-3}$ but can reach as high as $10^5$ during peak solar activity \citep{Stubbe1997}. Since the F layer is the highest deposition region of UV, it is the longest-lived and generally the least collisional of the ionospheric layers. Being the thickest and most dense layer of electrons, it generally has the largest impact on radio signals, especially for LWA observations of pulsars. The F layer is generally treated as consisting of two components, the lower F1 layer which only exists during the day, and the upper F2 layer which has the highest electron density and lasts through the night. The critical parameters for the F2 layer are the highest plasma frequency present, foF2 ($9000\,n_e^{0.5}$ Hz with $n_e$ measured in cm$^{-3}$), and the height of the corresponding layer, hmF2 (usually quoted in km).

\section{Observations}

\subsection{LWA1 Observations of Pulsar FR}   
 
\begin{figure}
\begin{center}
\includegraphics[width=.49\textwidth]{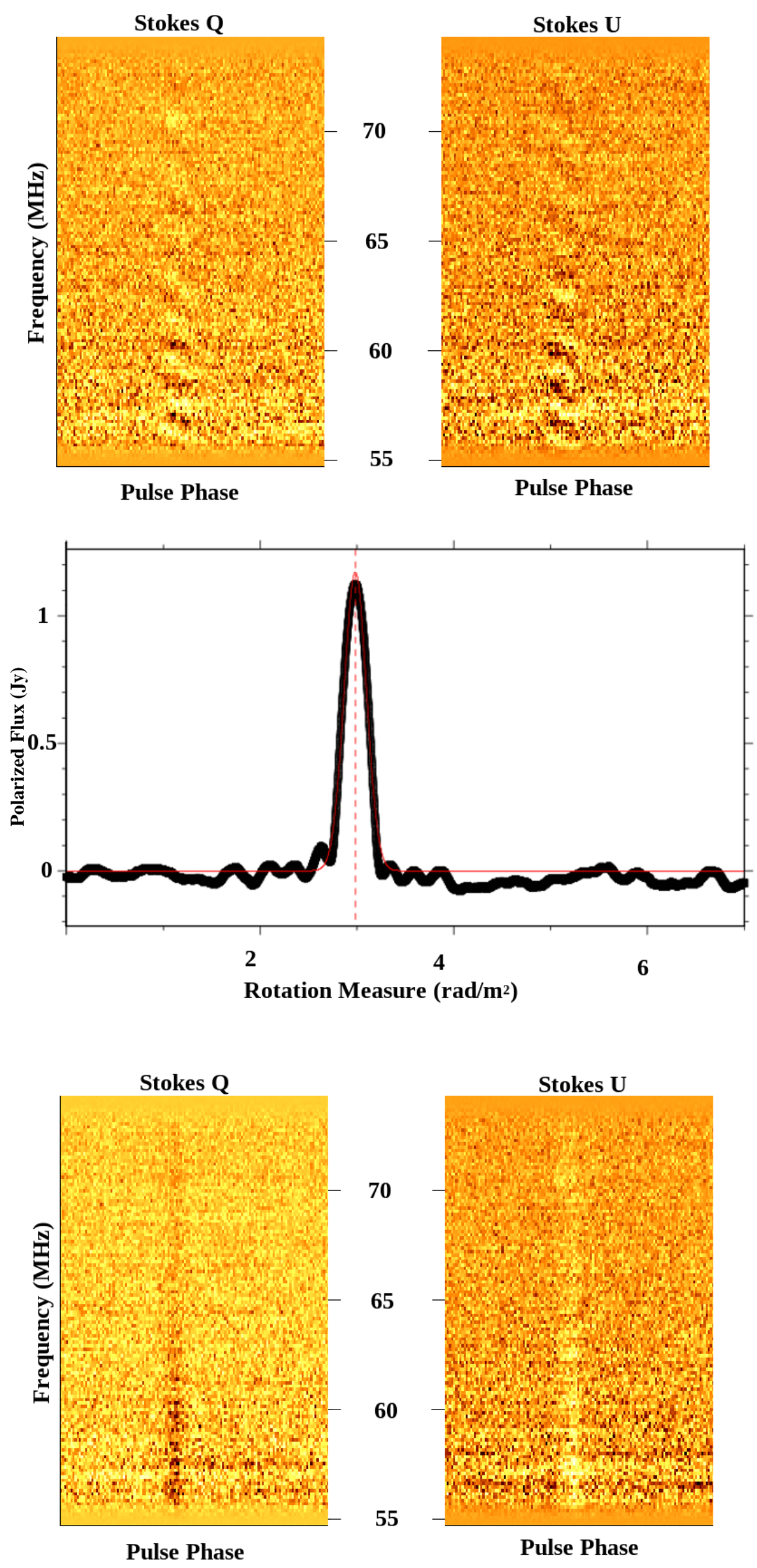}
\caption{\footnotesize (top) An example of FR seen in observations of the pulsar B0950+08. Each pulse phase is averaged and then shown verse their frequency, with bright spots representing bright signal. The diagonal banding seen in the center of the plots is a result of FR. Each frequency has slightly different linear polarization due to FR being frequency dependant (center) The program PSRCHIVE is used to test various RMs and see how that affects the polarization, Stokes U + Q. Near the true RM, the polarization spikes as frequencies constructively add. (Bottom) Applying this RM correction unwinds the FR, creating a plot with none of the banding seen above.\label{fig:Fday}} 
\end{center}
\end{figure}

This study uses the first station of the Long Wavelength Array (LWA1), co-located with the Jansky Very Large Array, to measure the amount of FR of nearby polarized pulsars. Pulsars are rapidly-rotating neutron stars left behind by the supernova explosions of massive stars \citep[e.g.,][]{Lattimer}. While the emission mechanism of pulsars is not well understood, it is believed that the emission is produced by relativistic charged particles trapped in the extremely strong magnetic fields of the neutron star. Often the magnetic axis is tilted with respect to the rotational axis, and occasionally the magnetic axis will point toward the Earth. Like a lighthouse, the magnetic field axis will sweep across the Earth as the pulsar spins, beaming a strong burst of radio emission at the Earth that produces the characteristic pulsed emission. Depending on the pulsar, this emission is often highly linearly polarized, providing valuable diagnostic sources for FR. Pulsar observations with LWA1 are described by \citet{Stovall1}, and LWA1 pulsar FR measurements are described by \citet{HSD16}. Since FR varies as $\lambda^2$, the large relative bandwidth, over 2 full octaves, or doublings of frequency, covering 20-88 MHZ, of LWA1 together with the large values of $\lambda$, 3-15 m, compared to LoFARs .125-.25 m wavelength for high band, that LWA1 measures result in unprecedented accuracy in measuring FR. This allows for an RM solution to be found with a limited amount of observing time, in this case only ten minutes of data is necessary where other instruments may need hours, and enabling us to track ionospheric effects. The effect of the ionosphere is isolated in these observations by observing pulsars well away from the Sun through sunrise, when changes in the ionospheric density profile due to solar radiation deposition dominate the observed time variations of the pulsar FR.

We obtained rotation measurements of the pulsar B0950+08 using LWA-1, a low frequency radio interferometer with a frequency range from 10 to 88 MHz \citep{Taylor,Ellingson}. The array consists of 256 dual polarization dipole antennas distributed over a diameter of roughly 100m. Pulsar observations are made using a beam-forming mode with a sampling rate of 19.8 MHz, resulting in a time resolution of 50 ns. LWA-1 pulsar observations are reduced using the LWA Software Library \citep{DWS12} and converted to a PSRCHIVE \citep{psrchive} format file for use. Once in the PSRCHIVE format, the RM of the pulsar is found using standard pulsar software to unwind the frequency variation of the plane of linear polarization. Figure~\ref{fig:Fday} has an example of the output of testing for various RMs.  Since each individual pulse is not bright enough to have its polarization measured, pulses are integrated in time to provide sufficient signal-to-noise in each polarization. B0950+08 is a bright pulsar with 85-90\% polarization in the LWA frequency band. The integration time needed to robustly measure the RM of this pulsar is about 10 minutes, and the average RM uncertainty for each 10 minute observation is approximately 0.03 rad/m$^2$.
    
The RM from the pulsar is caused by contributions from the material surrounding the pulsar, the interstellar medium, the solar wind and the ionosphere. The high degree of polarization of the radio emission indicates that the pulsar itself does not contribute a large RM, although there is no method to determine what it would be at this time. For the purpose of this study we assume it to be none. We further assume that the interstellar medium is changing at very slow rates, varying by hundredths of a rad/m$^2$ per year. The Sun-pulsar angle for these observations was around 60 degrees, i.e., the lines of sight do not pass close to the Sun, so the solar wind contribution should be small and steady. We assume that at these distances from the sun the solar wind does not change on the time scale of the observation and instead consider it a part of the ISM constant off-set. Thus the majority of the variation in RM over the course of the day should be due to the ionosphere, either from the density fluctuations or by the variation in line-of-sight component of the Earth's magnetic field. 

The interstellar medium provides a constant offset in the RM measurements. This absolute offset is generally not known for pulsars, although some attempts at determining the absolute offsets have been made. The current accepted value for the offset of B0950+8 can be found in \citet{Johnston}, with an absolute RM of -.66 +/- 0.04 rad/$m^2$, meaning we would add .66 to all of our measurements in order to correct for this intrinsic measure. We report on the absolute correction in this paper, which is the inverse of the pulsars absolute RM. We found this value to not correspond to our data, so to handle this issue, we allowed the RM offset to vary for each observation, attempting to reach a best fit for each model. We then compared the models and attempted to reconcile each model with the other by altering various parameters in the model.

The data used here were taken on 2016/09/23, 2016/10/05, and 2016/10/14, with each day consisting of multiple two-hour observations of B0950+08 separated by 15-minute buffer periods. Observations were taken in the morning beginning generally an hour before sunrise and continuing through to early afternoon. This was done in order to see the diurnal change in total electron content and the corresponding rise in RM.

\subsection{Ionograms}
\begin{figure*}
\begin{center}
\includegraphics[width=.95\textwidth]{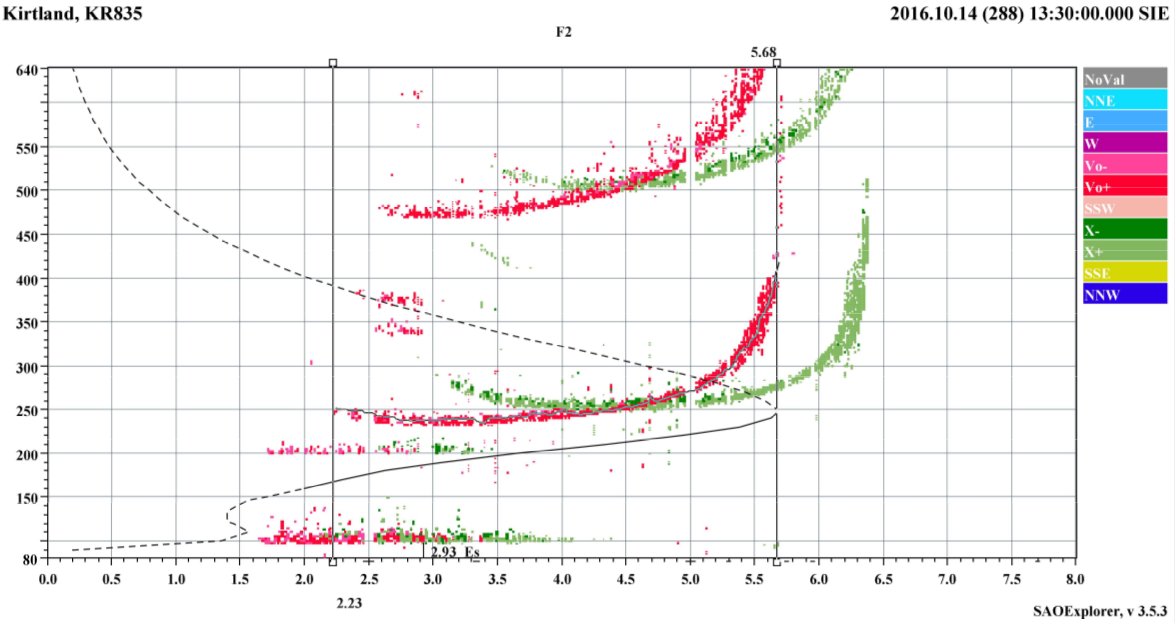}
\caption{\footnotesize An example ionogram from 2016 Oct 14, just after dawn at KAFB, a digisonde. The horizontal axis represents frequency, which can be translated to a density at which the signal reflects, and the vertical axis represents a psuedorange in km found by halving the travel time and assuming a vacuum medium. The red points represent o mode returns from the vertical direction, while the green points show x mode returns. Both single (lower) and double (upper) bounce traces are visible. The single-bounce signals are modified to account for the general group velocity delay all electromagnetic waves encounter in a plasma. The black line represents the fitted ionospheric density profile: solid portions of this line come from the measured o mode returns corrected for group delays, while the dashed portions represent modelled sections of the profile, including the Vary-Chap top-side mode above the F2 peak at 250 km. \label{fig:Ionogram}} 
\end{center}
\end{figure*}

The density profile of bottom-side of the ionosphere (i.e., below the F2 peak) was obtained using digisonde data. The Air Force Research Lab (AFRL) operates a DPS-4D digisonde, supplied by Lowell Digisonde International \citep {RGK08}, transmitting from Kirtland Air Force Base (KAFB) on the southern edge of Albuquerque, NM. Two digisonde receivers were used for this work, one co-located with the transmitter, and another located at the LWA-Sevilleta site about 80 km south of KAFB, providing redundancy in bottom-side profiles. Digisondes transmit a coded signal in short pulses, sweeping through frequency. The signal travels to the ionosphere where the ordinary-mode (``o mode'') polarization is reflected at the height where the plasma frequency in the ionosphere matches the transmitted frequency. The travel time of the return signal is measured, providing a height corresponding to that density. The frequency is typically swept from 1 to about 15 MHz, depending on the expected value of foF2. For this work digisonde sweeps were carried out every 3 minutes. The resulting frequency-height diagram is called an ionogram \citep{Mac91}: an example is shown in Figure~\ref{fig:Ionogram}. The o-mode trace is identified in the ionogram and is then fitted to a model of the ionosphere, providing a complete bottom-side density profile at a cadence high enough to track variations. The ionogram can only provide the bottom-side density profile, since vertically-transmitted signals at frequencies above the peak plasma frequency of the ionosphere pass through the ionosphere and escape without providing any reflected signal back to the ground. 

We compared the density profiles derived at the same times from the Sevilleta and Kirtland receive stations and found them to be in excellent agreement, with foF2 matching to better than 0.05 MHz and hmF2 matching to better than about 5 km. The reflection points for the two sites differ by about 40 km on the sky (half the horizontal separation of the sites), and we do not see clear evidence for a delay between the two sites at the 3-minute cadence of the measurements. We therefore regard these measurements as adequate representations of the ionosphere above LWA1, another 70 km away.

The unmeasured top-side density profile with height is generated using models developed from top-side ionograms obtained in the 1970s by NASA's ISIS-2 satellite \citep{Bil09}. The top-side profile is represented by a modified $\alpha$-Chapman function, known as a Vary-Chap function \citep{RNH07}. Using the peak height and peak density of the F2 layer derived from the ionogram as input, the top-side profile combines two functions, an F2 layer exponentially decaying with height with a particular height scale that merges into a plasmasphere with a power-law decay above it. This provides three degrees of freedom: the thickness scale of the F2 layer, the exponent of the power law decay and the transition height between the two profiles. All three variables have diurnal, seasonal and solar cycle dependencies, providing large amounts of freedom in the top-side shape of the ionosphere \citep[e.g.,][]{Nsumei}. A crude rule-of-thumb is that the top-side ionosphere contains about two-thirds of the TEC.

Digisonde measurements of the bottom-side density profile are available for the data taken on 2016/10/14. The 3-minute cadence ionograms were hand-scaled (i.e., the fundamental o mode signals on the ionograms were manually traced, rather than relying on the digisonde's automatic tracing software; e.g., Fig.~\ref{fig:Ionogram}) and then fitted to ionospheric profiles using the ARTIST 5 software package \citep{GKK08}, which is a Vary-Chap profile similar to that found in \citet{Nsumei}. The output is the electron density versus height in 2.5 km range increments. The bottom-side height profile is determined by the range-versus-frequency measurements of the digisonde, and the model top-side Vary-Chap profile is fitted onto the bottom-side profile by the ARTIST software as described above. With density profiles every 3 minutes, we can integrate (2), assuming that the ionosphere has a constant density profile over the visible sky, tracing the path of the pulsar across the sky and taking into account the changes in orientation with respect to the geomagnetic field.

\subsection{GPS Measurements of TEC}

\begin{figure*}
\begin{center}
\includegraphics[width=.95\textwidth]{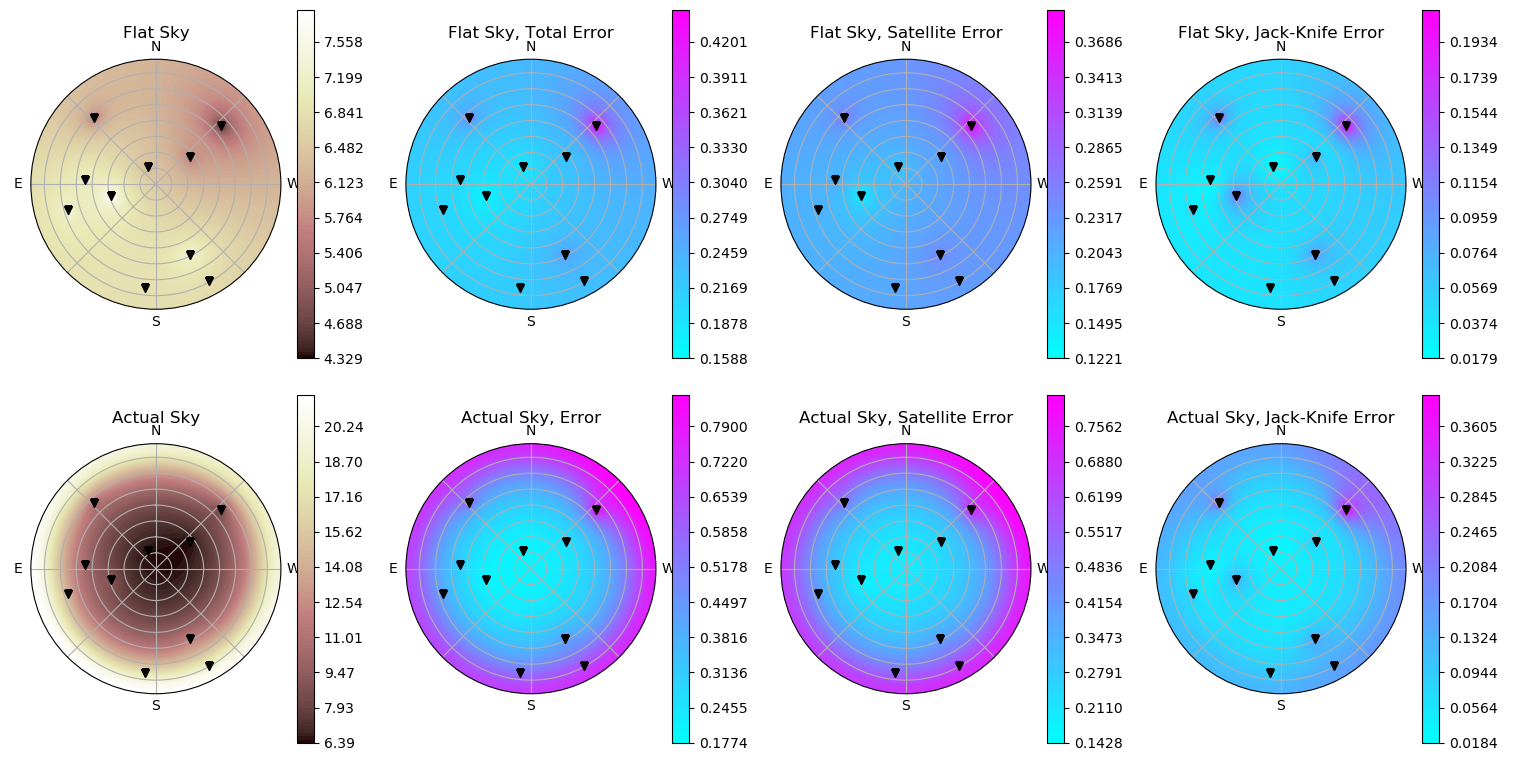}
\caption{\footnotesize The two plots are a map of the Total Electron Content (TEC) across the sky as seen above LWA1 at 1500 UTC on 2016/10/14. The black triangles represent the positions of the satellites. (top) Shows the model of the sky using only vertical TEC. This is to show how the satellites are converted to vertical TEC to model the points of the sky in between. (far left) A map showing the estimated TEC across the sky. (center left) A map of the total error of the TEC measurement. (center right) A map only showing the noise error of each satellite propogated onto each point in the sky, essentially showing temporal error in the satellite. (far right) A map showing only the jack-knife error for each point, displaying the estimated spacial error at each point  (bottom) Points are then converted back to slant TEC, which is a representation of the density as seen by an observer such as a radio telescope, with the appropriate error associated with each point. \label{fig:SkyMap} } 
\end{center}
\end{figure*}

GPS data can be used to measure the total electron content of the ionosphere by comparing the arrival times of GPS signals at two different frequencies. This study uses Novatel GSV 4004B dual-frequency receivers from the the Scintillation Network Decision Aid (SCINDA) program, supplied by the Air Force Research Laboratory (AFRL), to measure ionospheric TEC. This receiver is able to receive both the L1 and L2 GPS frequencies at 1575.4 MHz and 1227.6 MHz, respectively. The signals are transmitted with an embedded time code, which the receiver then decodes to determine the travel time of the signal on each frequency. These signals are converted to pseudoranges by multiplying the travel time by the speed of light in a vacuum and are recorded as P1 and P2, for the L1 and L2 pseudorange. The pseudorange differs from the true range by the group delay imposed on it by the ionosphere. These measurements are recorded every 10 seconds by the GPS receiver in Receiver Independent Exchange (RINEX) format, which is the standard format for GPS measurements. The difference in the pseudoranges can be converted to total electron content using the formula $TEC_{P} = 9.5196 \left(P1 - P2\right)$ \citep{Hernandez}, where P1 and P2 are measured in meters and TEC is reported in Total Electron Content Units (1 TECU $= 10^{16}$ electrons/m$^2$). Due largely to multipath effects, signals from satellites low on the horizon are suppressed using a hyperbolic tangent function centered at zenith angle 65 degrees. Multipath refers to the reception of signals from the same satellite arriving from different directions, with different delays, primarily due to reflections off nearby objects. In addition to the pseudoranges, a delay can be found from the phase of the signals themselves, by converting the phase of one signal to the other, and using the formula TEC$_{\phi} = 1.812 (L1 - L2 \times 1.5745/1.22760)$. However, there is 2$\pi$ ambiguity when using only phases. The solution is to use the phase-referenced TEC measurement and calibrate the signal to the pseudorange TEC measurement, which uses the formula TEC$ = \textrm{TEC}_{\phi} + \left|\textrm{TEC}_{P} - \textrm{TEC}_{\phi}\right|_{arc}$, where the brackets indicate the average of the TEC measurements over a phase-connected arc. This measurement represents the TEC along the line of sight from the satellite to the receiver, and is known as the slant TEC (sTEC). The noise level of this measurement is typically .1-.4 TECU, determined by taking a 5 minute average in signal and computing the variance. This error is recorded as satellite error in Figure~\ref{fig:SkyMap}. The measurement that we are interested in is the vertical TEC (vTEC), which is a representation of what the TEC would be if the satellite were at the zenith. To convert sTEC to vTEC, we follow the usual procedure of collapsing all electrons to a single height of 350 km, known as the thin-shell model. Using the law of sines, the vTEC of a given measurement is found using the technique described in \citet{Sotomayor}.

\begin{figure}
\begin{center}
\includegraphics[width=.49\textwidth]{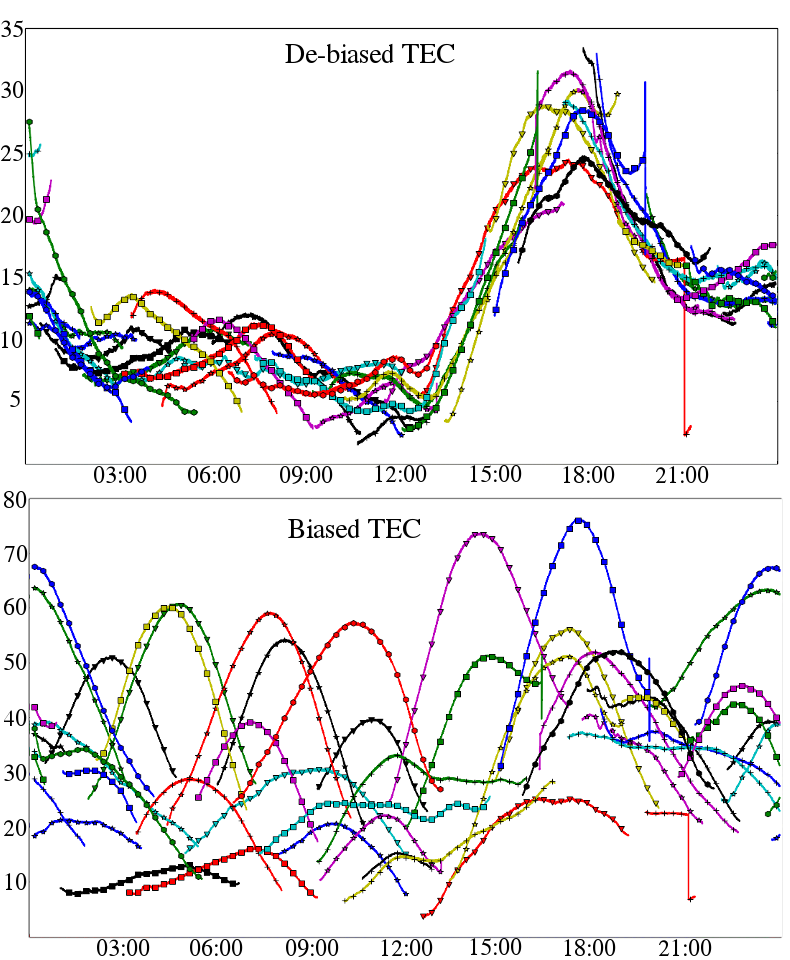}
\caption{\footnotesize (top)  A plot of the vertical total electron content (vTEC) for each satellite derived from the data in the bottom panel, after correction for elevation and the bias of each satellite and receiver. All times are in UTC. (bottom) Slant TEC data from GPS satellites visible at the LWA1 site over 24 hours on 2016/10/14, measured with SCINDA receivers. These data also carry the satellite-receiver biasses. \label{fig:vTEC}} 
\end{center}
\end{figure}

Each measurement still contains an offset from the true TEC that is a combination of electronic delays in the individual GPS receivers as well as relativity corrections, tropospheric effects and nonlinear ionospheric effects. As a result, there is a bias associated with each satellite and each receiver. To solve for this bias we used a method similar to the Kalman filtering technique discussed in \citet{Carrano}. We can exploit the fact that the biasses are essentially constant in space and time for each satellite-receiver pair, and that vTEC across the sky should be relatively flat. This allows us to find each satellite bias as a constant correction applied to the sTEC value, and the corresponding corrected vTEC tracks should match each other. The error between the satellite and each other satellite is minimized over several arcs over the course of 24 hours, since we expect any satellite errors to be systematically random and not correlated with a particular point in time or space. The positions of the GPS satellites are found using the package PyEphem \citep{Rhodes} and the Two Line Element (TLE) sets available through Space-Track.org. An example of the output of both the biased measurements and the de-biased vTEC values can be seen in Figure~\ref{fig:vTEC}. Note that once the outputs are de-biased, they follow the expected daily cycle of ionospheric density. In addition to modelling and de-baising the satellites we also model and subtract away the plasmasphere using the method used by \citet{}. This model provides an empirically derived model of plasmasphere data by \citet{Carrano} which uses sun spot number, day of the year, and McIlwain \textit{L} parameter to in order to model the the inner plasmasphere as well as an empirical model of the plasma trough and outer plasmasphere. These electrons represent the contribution from the total plasmasphere, which should be discounted from FR calculations since this plasma is sufficently far from the magnetic field to not impart any significant FR on the signal.

A model of vTEC across the sky is made from the series of de-biased TEC measurements from each satellite. The model uses a linear weighting to estimate the total electron content at each point on the sky from the vTEC of each visible satellite, and the weighting based on the angular distance to the point from each satellite:
	
\begin{gather}
\textrm{TEC} = \sum_i \mu f_a \frac{\textrm{TEC}_i}{d_i} \\
\mu = \frac{1}{\sum_i \frac{f_a}{d_i}} \\
f_a = \frac{1}{2} \left[1 - \tanh \left(12 \left(\phi - 65 \right) \right)\right]
\label{eq:htan}
\end{gather}

\noindent where $\mu$ is the normalization factor to ensure the weighting equals unity. $f_a$ is a weighting function that down-weights satellites at elevations below 30 degrees, and $\phi$ is the zenith angle of the satellite. $d_i$ is the angular distance between the calculated point and the $i$-th satellite, calculated using the Haversine formulas. At low elevations GPS signals tend to have much stronger multipath effects, and their TEC measurements quickly lose accuracy, so signals are down-weighted according to equation (\ref{eq:htan}). The vertical TEC map is then converted back into slant TEC. The error is estimated for each satellite and then is combined in quadrature with a jack-knife re-sampling to find the spatial error at each point. To find the jack-knife error, each satellite is removed one at a time and the map recalculated. This provides a range of values at each point from each satellite, and a standard deviation from this is used as the spatial error. An example of a sky model can be seen in Figure~\ref{fig:SkyMap}, with both the vTEC model and the sTEC model given for comparison. This model is then used to predict the RM contribution of the ionosphere at a given location on the sky, discussed further in section \ref{sct:RMP}.

\section{Rotation Measure Predictions}
\label{sct:RMP}
We developed two models for predicting ionospheric contribution to RM. The first uses the GPS total electron content model together with the International Geomagnetic Reference Field 12 (IGRF-12). Essentially the entire column of electrons in the ionosphere along the line of site is collapsed to a single point. Deriving the magnetic field at this point from IGRF-12, we produce the expected RM contribution from the ionosphere:

\begin{gather}
\textrm{RM} = 2.62 \times 10^{-13} \left( TEC \cdot B_{IPP} \right)
\end{gather}

\noindent where $TEC$ is the ionospheric total electron content along the line of sight to the pulsar and $B_{IPP}$ is the magnetic field at the ionospheric pierce point. This model assumes that there is little structure between satellites, as well as a relatively even distribution of electrons in the profile of the ionosphere. Essentially, by collapsing the electrons to a layer of a single height we are altering the contributions of the electrons to the RM. Electrons below the chosen height, where $B$ is stronger than assumed, are down-weighted while electrons above the height are effectively given higher weight. This resulted in an intrinsic RM measurement for pulsar B0950+08 of 1.72 +/- 0.075. Here we use the RMS residual difference between the LWA observations and the model to estimate the error of the measurement. 

The second method for determining the RM is to use to use the digisonde electron density profile to determine the ionospheric contribution. The integration in (2) is carried out with $B_\parallel$ calculated along the line of sight of the pulsar using IGRF-12. This model requires that there is no horizontal structure in the ionosphere, since we are using the same density profile at all locations. As well it requires that the top-side model is correct since the majority of the RM contribution is expected to be from that region. The intrinsic RM measurement for this pulsar using this method is estimated to be 2.04 +/- 0.091.

\begin{figure}
\begin{center}
\includegraphics[width=.49\textwidth]{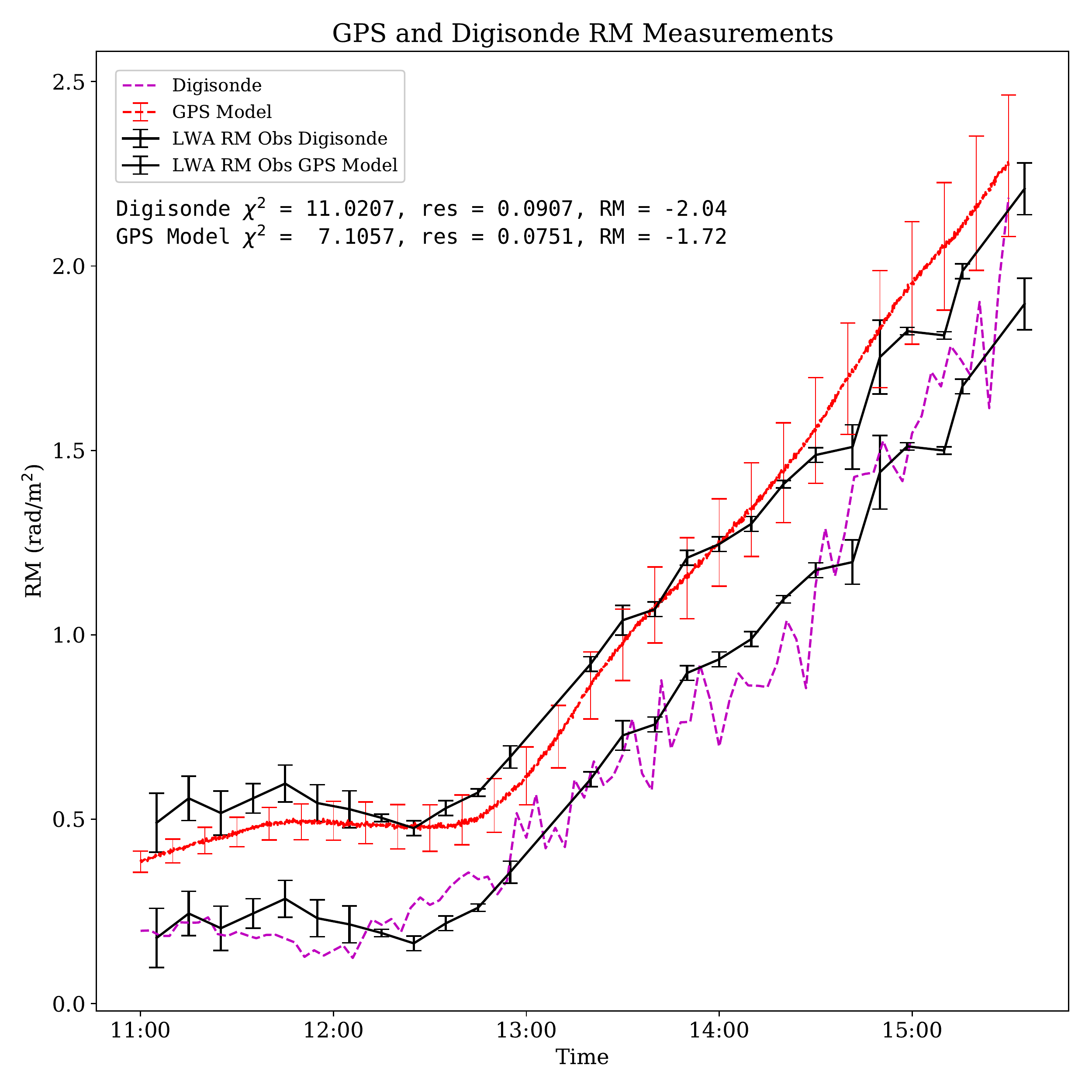}
\caption{\footnotesize Plots of data taken on Oct 14, 2016 of the predicted RM using the GPS model (red dashed line) and the Digisonde model (purple dashed line). The black lines represent the LWA1 RM observations with a constant offset determined by minimizing the $\chi^2$ associated with the offset. Also shown is the $\chi^2$ between each model and the LWA1 RM measurements, as well as the average residual RM after subtracting the model from the measurements. The large difference in the RM offset can be attributed to errors in the models. \label{fig:GPSDigiRM}} 
\end{center}
\end{figure}

The results from these models can be found in Figure~\ref{fig:GPSDigiRM}. Each model, shown with dashed lines, is compared to the LWA1 RM measurements, shown in solid black lines. The reduced $\chi^2$ was calculated for each model as well as the average residual RM after subtracting the model from the measurements. The residual RM can be used to indicate the accuracy of the model. If the residual RM were zero, then we could say that the model effectively removes the ionosphere. A constant offset in the RM was determined for each model, representing the unknown interstellar contribution to the measured RM. This constant offset was found by fitting for the RM offset with the lowest $\chi^2$. With the $\chi^2$ values being small and the residual RM for each model being far less than the difference in RM offset, the error between the two must result from an error in one or both of the models. The following sections will attempt to reconcile the differences between the two models with respect to RM offset.

\subsection{Correcting GPS Predictions}
\label{ssct:GPS2D}
We began by making the assumption that the digisonde measurements were accurate and that we need to find a way to correct the GPS model to match the RM offset found from the digisonde model density profile. There are two methods to alter the thin-shell GPS model to fit the offset provided by the digisonde model. The first assumes that the GPS model over-predicts the number of electrons in the ionosphere due to some unforseen delay. This would manifest as a delay between the signals and would be invariant in space and time. This seems extremely unlikely, since this type of delay would be identified in the bias correction used to create the TEC model. If this delay varied in space and time, it would have to do so at exactly the rate each satellite passes over our GPS receiver, and do so at a constant rate each day. This seems extremely unlikely, so we dismiss this possibility.

\begin{figure}
\begin{center}
\includegraphics[width=.49\textwidth]{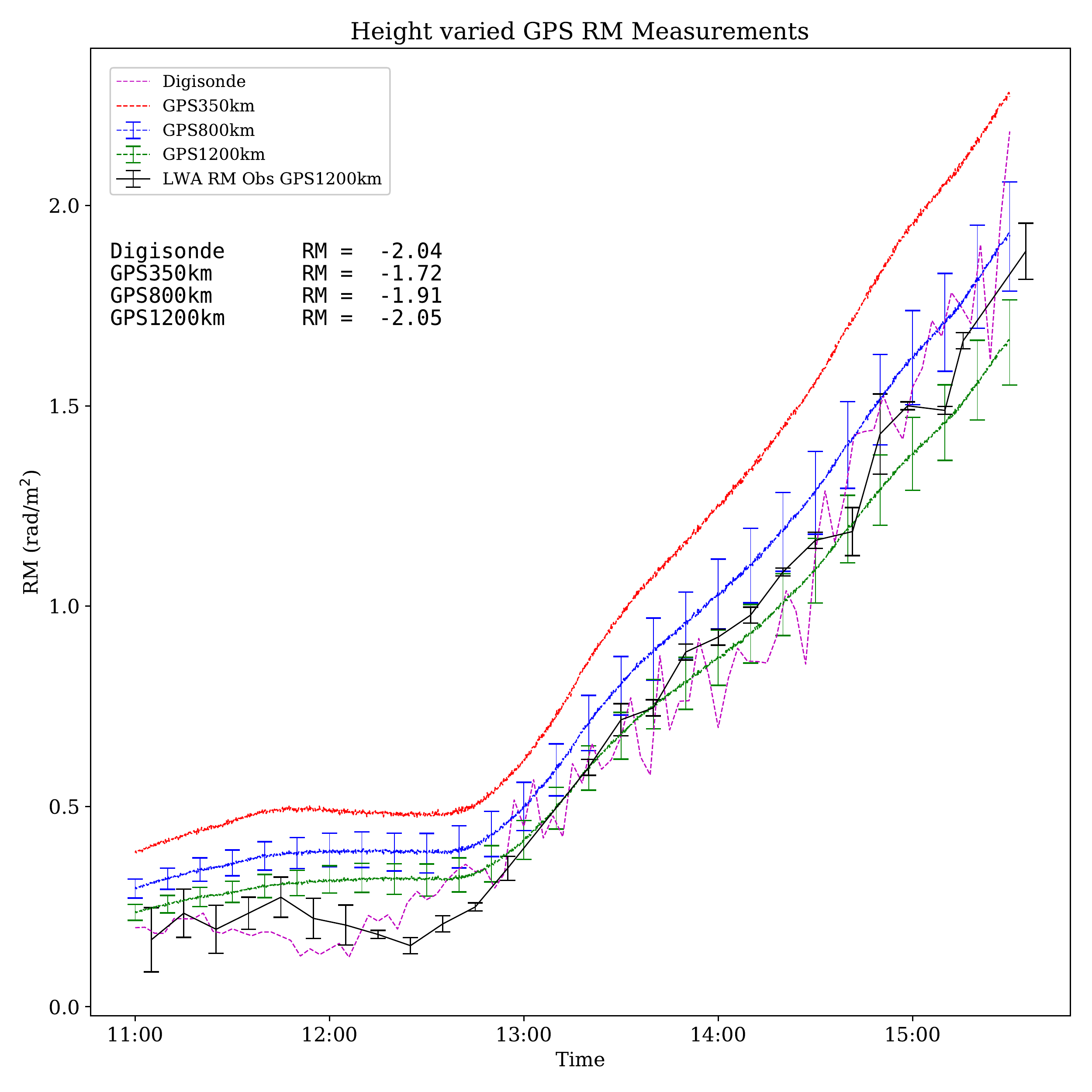}
\caption{\footnotesize Plot similar to Figure~\ref{fig:GPSDigiRM} with added plots with GPS model using 1000 km and 1400 km as the height of the thin shell. Note that the RM offset for the 1400 km shell is approximately the same as that of the digisonde model. \label{fig:GPSHeightChange}} 
\end{center}
\end{figure}

The second option is that the height chosen for the thin shell is incorrect. The height represents the median contribution height, such that the electrons below the chosen height contribute exactly the same as those above. By raising the height, all of the electrons contribute less to the overall RM. This also implies that the median density (the point where equal numbers of electrons lie above and below the point) is above this shell height, since the magnetic field up-weights electrons lower in the atmosphere. If we do this, we find that the shell height would need to be 1400 km, as seen in Figure~\ref{fig:GPSHeightChange}, in order to produce a RM offset similar to that found by the digisonde. This is implausible, since the F2 peak electron density is near 300 km height.

\subsection{Correcting Digisonde Predictions}

\begin{figure}
\begin{center}
\includegraphics[width=.49\textwidth]{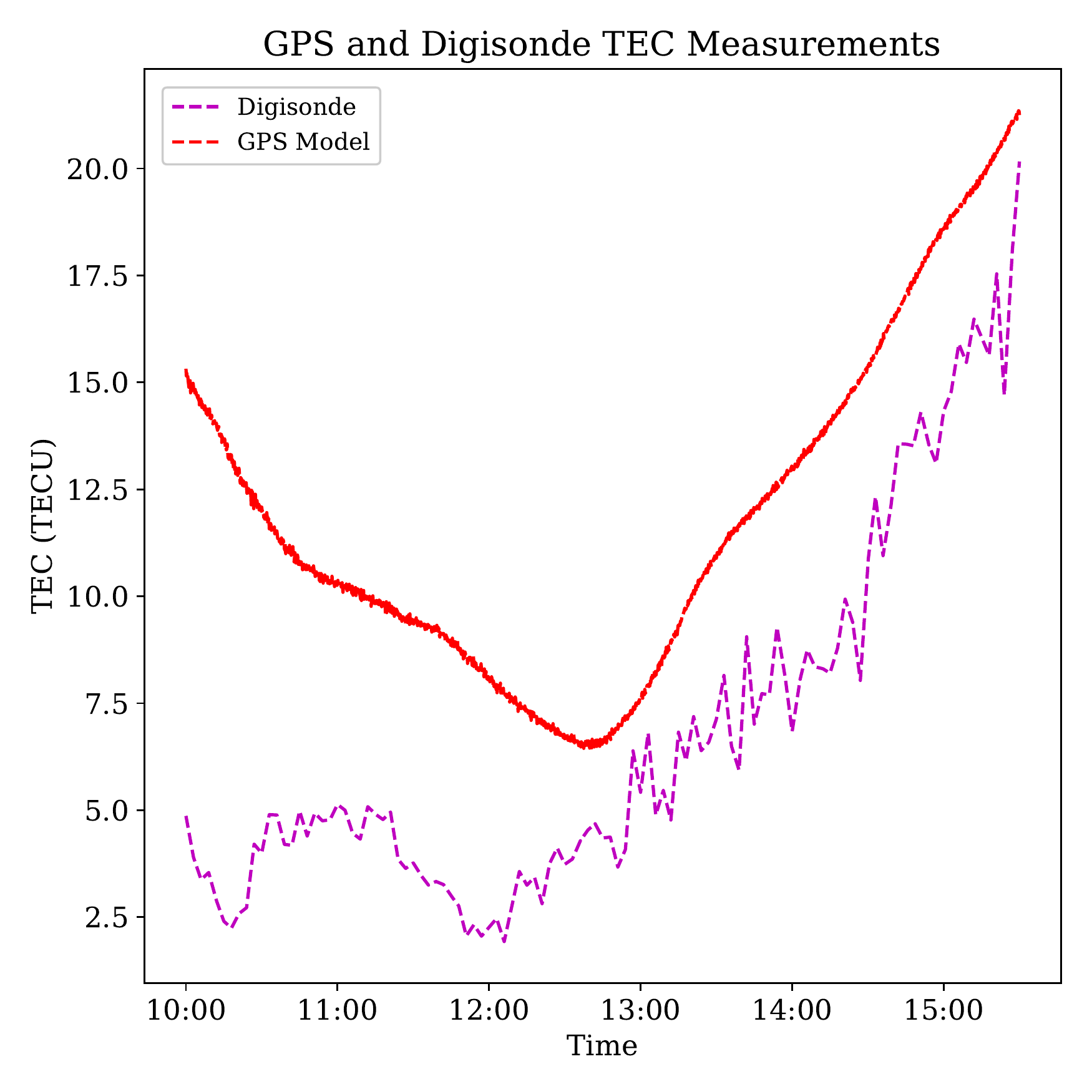}
\caption{\footnotesize The integrated electron density of the predicted profile from digisonde measurement compared to the measured slant TEC from GPS data as it looks toward the pulsar. This is much higher toward morning due to the low angle looking at the pulsar. CLearly shown here are the significant difference in TEC comparing the digisonde and the GPS \label{fig:GPSvIono}}
\end{center}
\end{figure}

If on the other hand we assume that the GPS RM offset is correct, we find that the digisonde under-predicts the RM contribution of the ionosphere. Since the electron densities of the bottom side are directly measured, we will take these as being accurate. The largest component of ionospheric RM is from the digisonde profile above the F2 layer, which is modelled but not measured, suggesting that a solution is to modify the top-side profile to account for the difference in RM. By integrating the digisonde electron density profile we find that the the TEC in the digisonde profile is always well below the GPS measurements. We therefore infer that the ionogram profiles do not account for a large fraction of the ionospheric TEC. Figure \ref{fig:GPSvIono} compares the GPS TEC data (red curve) with the TEC derived by integrating through the digisonde profiles (blue curve) on 2016 Oct 14: the ionogram profiles underestimate the total TEC by 50\% during nighttime hours and about 30\% during daylight. In order to increase the amount of RM provided by the digisonde model, we need to increase the number of electrons in the top-side model. We need the electron density to be below the peak density, or we would see them in the ionogram provided by the digisonde, and we need them to be close to the peak layer in order to contribute meaningfully to the RM. It is desirable to maintain a Vary-Chap profile for the top side, since this was based on observations. However, electron density profiles provided by incoherent backscatter radars, such as Arecibo, suggest that the top of the F2 peak is smoother than the Vary-Chap model implementation allows \citep{Morton}, indicating that the F2 layer provided by the digisonde profiles is too narrow, and that there is likely an extended region of electrons not accounted for in the peak F2 region by the Vary-Chap profile.

In order to raise the RM of the digisonde profile, we attempted to increase the number of electrons in the profile to that found in the GPS TEC measurements. The model for the top-side profile can be found in \citet{Nsumei}, and it consists of a hyperbolic tangent function to represent both the F-layer top side as well as a power law decay for the transition to the plasmasphere. The top-side model uses two parameters derived from the ionograms as inputs: the height at which the peak electron density occurs (hmF2), known as the scale height, and the peak electron density (NmF2, i.e., the density corresponding to a plasma frequency of foF2). This leaves three parameters not determined by the bottom-side profile: the thickness of the F-layer, the height where the transition between the ionosphere and plasmasphere occurs, and the power law decay exponent used to represent the plasmasphere. Each of these parameters has a known possible range based on the time of year, time of day and magnetic latitude of measurement. As such we only looked at values for the profiles that were within the accepted ranges when altering the profile. We tried to determine a set of parameters that gave density profiles that matched the TEC value measured by GPS. This provides a two dimensional surface of acceptable values within the parameter space. We can then find the RM each profile on the surface gives when combined with the magnetic field information. The RM for all of the points on the surface vary by only several thousandths of a rad/m$^2$. Since our only value measured was RM, we took all of these values as possible solutions, and used the output RM measure for our adjusted model. Often, a set of acceptable values within the physical parameter space defined in \citet{Nsumei} could not be found. This is taken as a failure of the models to produce physical results and we do not consider these further in the analysis.

\begin{figure}
\begin{center}
\includegraphics[width=.49\textwidth]{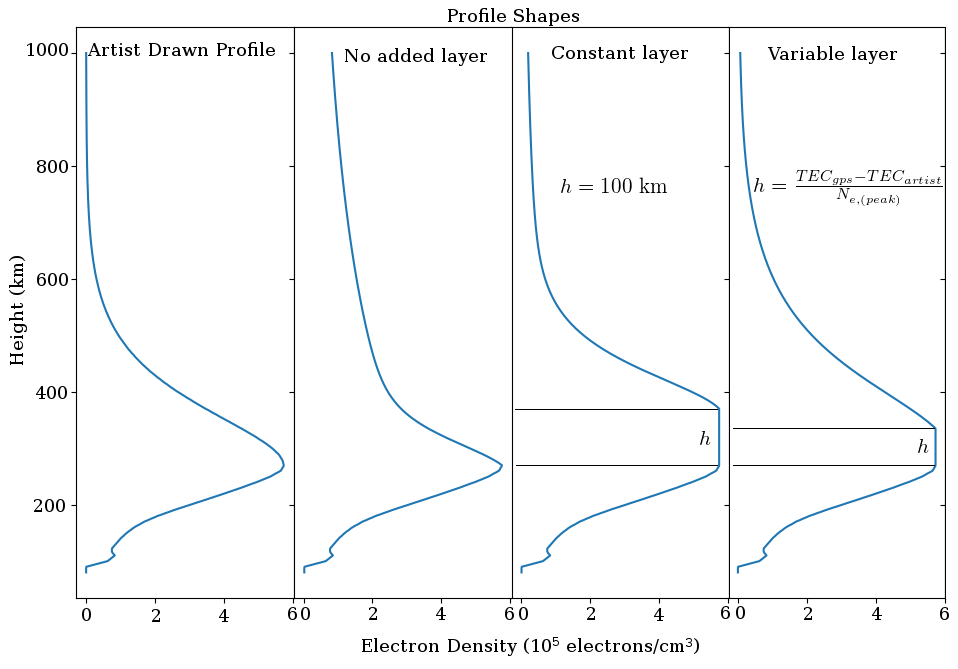}
\caption{\footnotesize Representative electron density profiles incorporating changes to the top side ionosphere template. (far left) Unmodified Vary-Chap profile provided by ARTIST 5. (center left) The parameters for the Vary-Chap profile are modified without an extra layer to ensure that the integrated TEC is equal to the GPS TEC. This method generally fails to produce physical parameters e.g. having an F-layer thickness that is above the transition height between the F-layer and the plasmasphere. (center right) a constant 100-km thick layer with density equal to the peak density is added above the F2 peak, with the height of the upper edge of this layer then used as thebase of a standard Vary-Chap profile. The Vary-Chap profile is solved for parameters that return an integrated profile equal to the GPS TEC (far right) a layer with density equal to the peak density is added above the F2 peak, with the thickness set to provide the difference between the GPS calculated TEC and the integrated TEC from the original ARTIST profile. The top of the layer becomes the input height to the Vary-Chap profile and again is solved to ensure that the profile-integrated TEC is equal to the GPS TEC. \label{fig:ProfileShape}}
\end{center}
\end{figure}

We examined three different approaches to increase the number of electrons in the top-side model, demonstrated in Figure~\ref{fig:ProfileShape}. The first technique fits the Vary-Chap profile so that the digisonde-predicted TEC is forced to match the GPS-provided TEC. However, a large portion of the data set was unable to produce a valid set of parameters. In general, solutions would return values that had extremely low power-law exponentials, an extremely high transition height and an extremely large F layer. 

In order to find a model that produced acceptable parameters a layer of electrons was added just above and equal to the peak density layer. This represents a layer of electrons that top and bottom side sounders may not be able to see since they are very close to the peak height. The first model adds a 100 km-thick layer just above the ionogram-predicted peak height hmF2, filled with the density corresponding to NmF2. Above this layer a Vary-Chap top-side profile is added using the top of the 100-km layer as the peak F2 height, and the total predicted TEC of the profile is set equal to that provided by GPS measurements. This technique provided parameters for the Vary-Chap top-side profile that were closer to values discussed as typical in \citet{Nsumei}, and therefore the profile could work as a model for the ionosphere. 

The third technique used the difference between the ionogram-derived TEC and the GPS-measured TEC to modify the thickness of the peak height layer. Again, above the added region of electrons a Vary-Chap profile was provided using the height of the top of the layer and the GPS TEC as inputs. Again, this approach provided Vary-Chap parameters much closer to those described in the Nsumei models, suggesting that it may be acceptable and physical.

\begin{figure}
\begin{center}
\includegraphics[width=.49\textwidth]{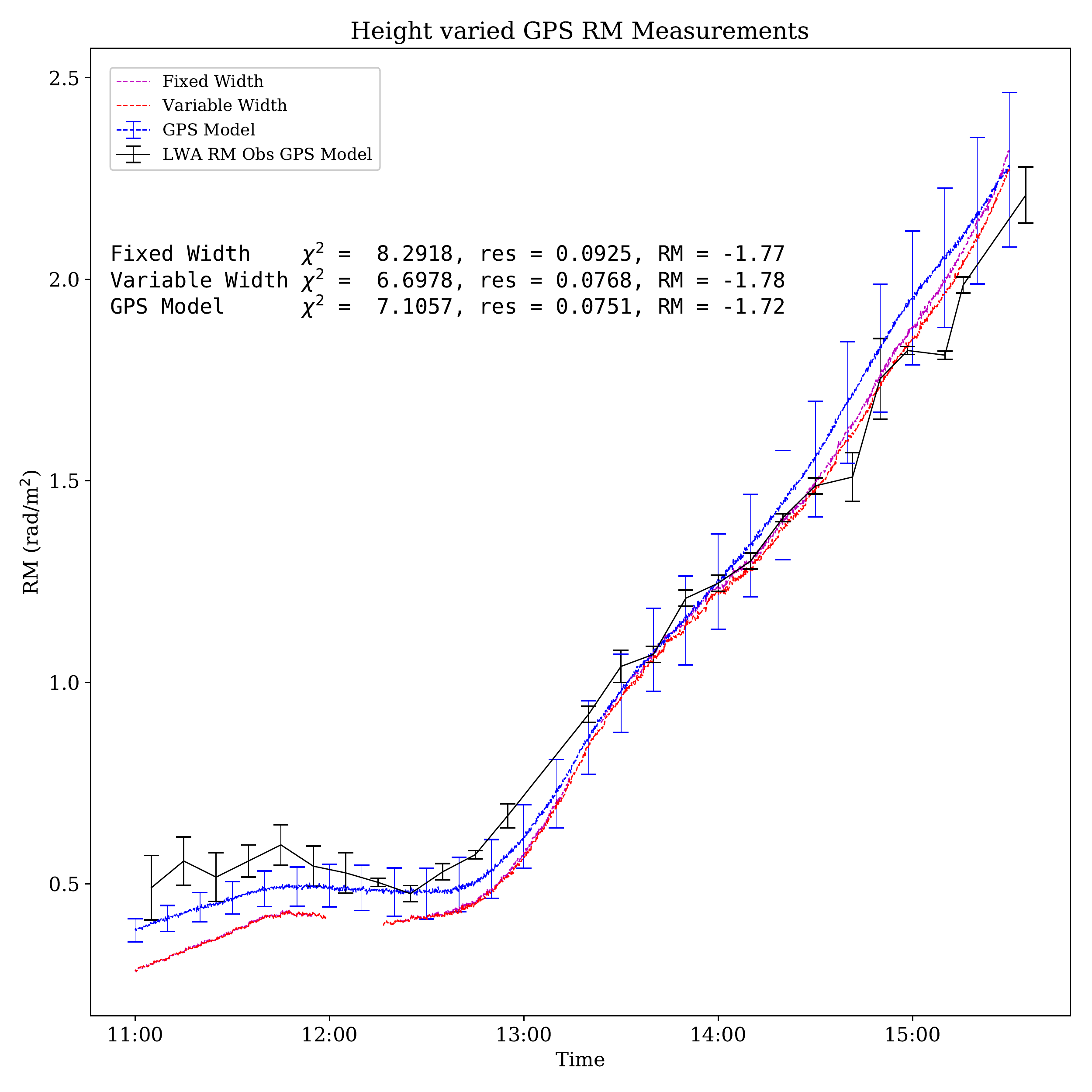}
\caption{\footnotesize Thin-shell GPS model compared to the digisonde model corrected in two ways. The first model correction is the fixed thickness, where a 100 km layer of peak height thickness was added just above the F-layer, and scaled thickness profile, where a layer of peak height thickness was added equal to the difference in TEC between the GPS model and the integrated digisonde model. The $\chi^2$ and residual RMs are all comparable to each other, indicating that the models are within good agreement.  \label{fig:VaryChapCorr}} 
\end{center}
\end{figure}

\begin{figure*}
\begin{center}
\includegraphics[width=.95\textwidth]{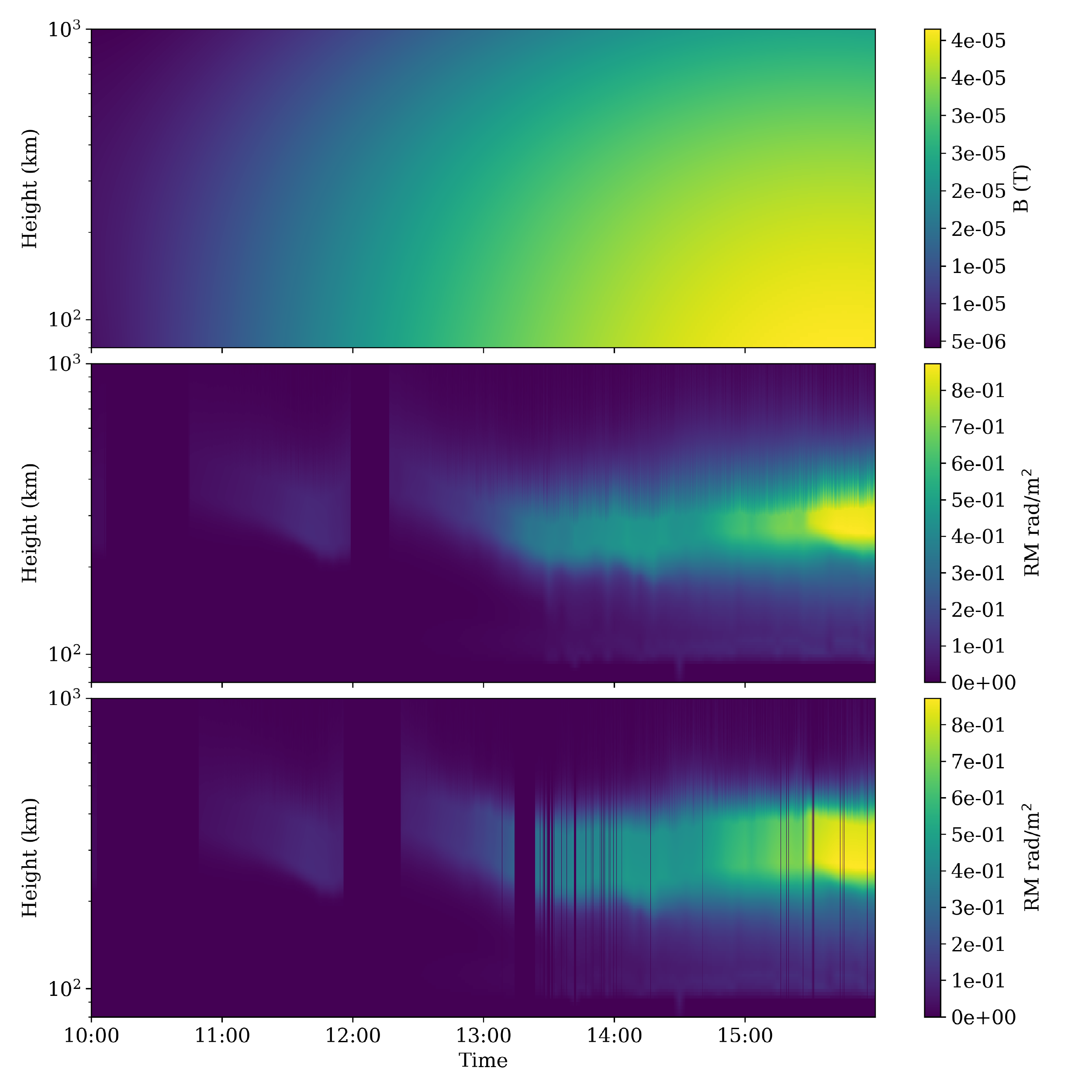}
\caption{\footnotesize (top) Magnetic field component along the line of sight from LWA-1 to the pulsar B0950+08 as a function of height and time on 2016/10/14. All times are in UTC. The observation begins with the pulsar at an elevation of 10 degrees and looking almost due east. The pulsar reaches a peak elevation of 63 degrees looking directly south at ~15:30 UTC. The magnetic field data are taken from the International Geomagnetic Reference Field 12. (middle) The RM contribution as a function of height and time in the model in which a constant-thickness layer of density equal to the F2-peak density is added at the F2 peak in the ionogram profile, where the thickness is set so that the profile matches the GPS-derived TEC. Blank spots are times in which there was no Vary-Chap profile was able to fit all of the parameters. (bottom) As above, a RM contribution by height, but with a profile generated by adding a layer of peak electron density with width sufficient to make up the difference between the GPS-measured TEC and the TEC derived by integrating the ionogram model profile. Again, blank spots represent times where no Vary-Chap profile was available to fit the given input. \label{fig:MagField}} 
\end{center}
\end{figure*}

While the first model was unable to find acceptable parameters for most of the measurements throughout the day, the results of the modifications from the second and third models can be seen in Figure~\ref{fig:VaryChapCorr}. These models produce RM values that are very close to one another, with the offsets attributed to the ISM being nearly identical. Essentially, by adding this extra layer of electrons into the model, we are forcing a thin shell to exist in the ionosphere. This becomes even more apparent when we examine the RM contribution as a function of height seen in Figure~\ref{fig:MagField}. The top plot shows the dot product of the magnetic field with the line of sight to the pulsar as a function of height and time. This function is then combined with the electron density distribution and integrated to produce the model RM. The middle plot shows the RM contribution for the constant thickness model as a function of height and The bottom plot shows variable thickness model. In both RM contribution plots, the (constant) 350 km thin-shell height is plotted as a solid black line while the digisonde-inferred hmF2 height is plotted as a thin blue line. This illustrates that essentially a shell of electrons is created by this model just under the 350 km thin-shell height.

There is currently at this time not enough data in the mid magnetic latitudes for determining whether the added layers of near peak density are physical or not. Obviously a real electron density profile will be a smooth function without any constant section, but these models provide a good approximation of plausible models. Comparing these model electron density profiles with electron density profiles derived from Arecibo incoherent-backscatter radar data \citep{Eccles}, it is not implausible that this type of electron density would exist. Combining this with the good agreement between GPS and digisonde measurements for RM correction suggests that this is the more plausible solution for reconciling disagreement between the two models, rather than altering the GPS thin-shell height.

\section{Comparison with global TEC models}

\begin{figure*}
\begin{center}
\includegraphics[width=.95\textwidth]{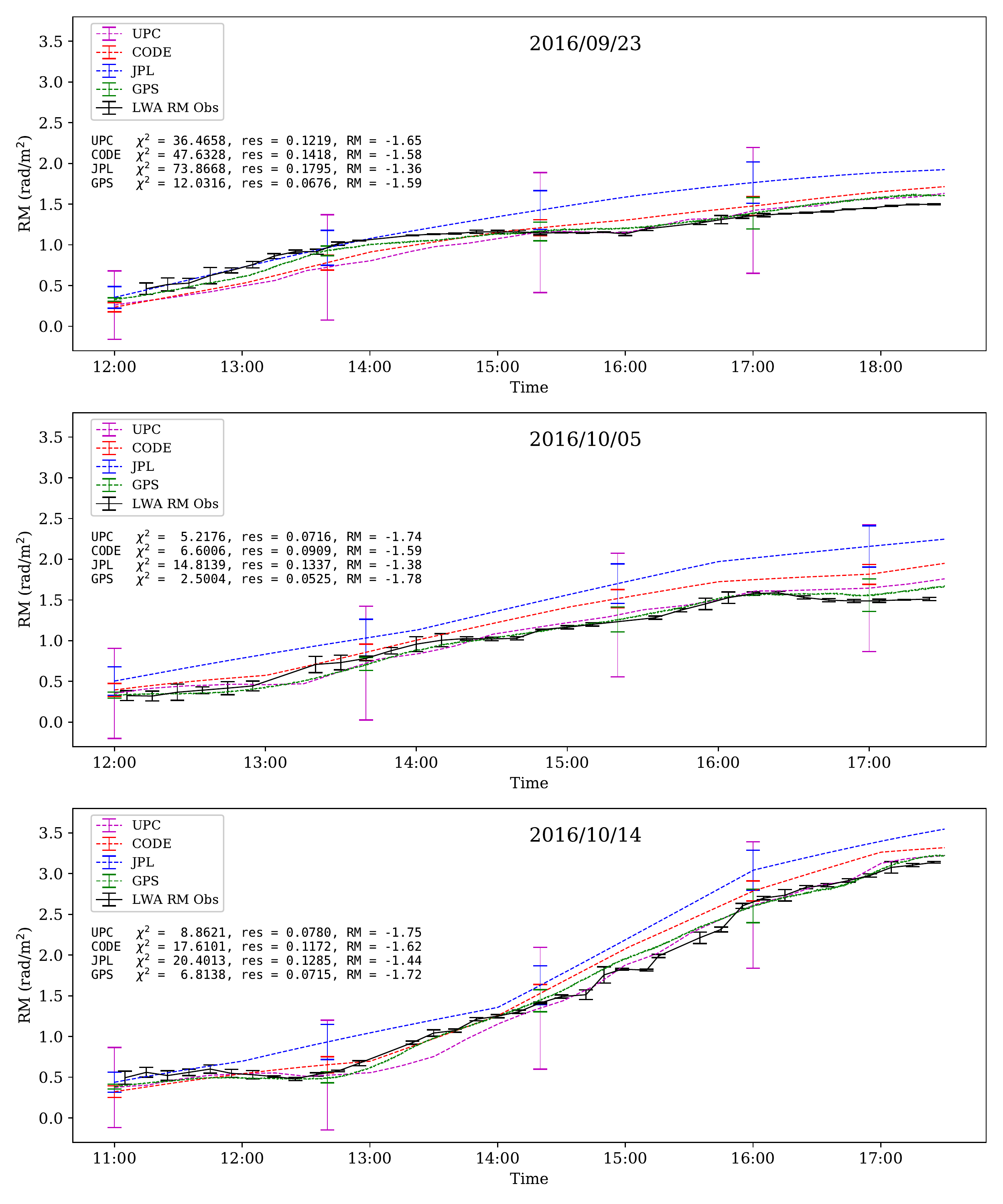}
\caption{ \footnotesize RMs of the pulsar B0950+08 as a function of time (marked with the dotted line in black and plotted with error bars) seen from the LWA station 1 site on 2016/09/23, 2016/10/05, and 2016/10/14. The observations begin before sunrise and continue until just after local noon. The data are compared with three different models: the GPS measurement prediction and model (marked in green), NASA JPL IONEX global sky model (Blue), CODE IONEX global sky model (Red) and the gAGE UPC 15 minute rapid solution model (Purple). \label{fig:161014TS}}
\end{center}
\end{figure*}

Interpolation within global GPS TEC models is commonly used to determine TEC at a given location and time. In Figure \ref{fig:161014TS} we compare the thin-shell results using GPS TEC measurements at the observatory with three standard global TEC models derived by assimilating GPS TEC measurements onto a global grid and then interpolating to the location of LWA1. The three models are the Jet Propulsion Laboratory (JPL) model, the Polytechnical University of Catalonia (UPC) rapid 15 minute solution, and the Center for Orbit Determination in Europe (CODE) model. The JPL model [https://iono.jpl.nasa.gov/] assumes a 3-shell ionosphere with slabs centered at heights of 250, 450 and 800 km, and updates every 15 minutes using $\sim$ 200 GPS stations on a grid with a spatial resolution of order 5 degrees, however these 15 minute solutions are compressed into 1 hour solutions when transmitted. The UPC model [http://www.gage.upc.edu/drupal6/forum/global-ionospheric-maps-ionex] uses a similar array of GPSs, but provides a 15 minute interval of solutions, but with larger errors. The CODE model [http://aiuws.unibe.ch/ionosphere/] employs a spherical-harmonic expansion with 2-hour cadence. TEC values were linearly interpolated between grid points as well as in time for all three models. We plot the results for all three dates on which pulsar measurements of ionospheric FR were made, and in all cases the thin-shell model using TEC data from the observatory out-performs the JPL, CODE, and UPC profiles. There are likely two reasons for the superior perfromance of the local TEC measurements: the time cadence of solutions is much higher, providing detailed data every 10 seconds rather than 15 minutes to 2 hours depending on global model; and by co-locating with the instrument, the lines of sight from satellite to receiver are virtually identical to that seen by the telescope, providing a more accurate picture of the ionosphere.

\section{Conclusion}

We have used measurements of FR in the linearly-polarized signal from a radio pulsar starting before sunrise and continuing well into daylight in order to determine the ionospheric contribution to FR. We compare these measurements with several ionospheric models and compare the accuracy of these models to that of the pulsar measurements. The new feature of this study is the use of a digisonde to measure the bottom-side density profile of the ionosphere with 3-minute cadence during the pulsar observations as well as TEC measurements using a GPS receiver located at the observatory.

We find that the standard ionospheric electron density profiles resulting from fitting the ionograms cannot reproduce the RM measurements, and have investigated ways to account for the differences between the models and the measured RM. Based on the fact that the GPS TEC data and the bottom-side ionospheric density profiles are both direct measurements, we conclude that the top-side density profile modelled by ARTIST 5 from the bottom-side parameters is incorrect. We therefore investigate modifications of the top-side profile that are physically plausible and match the TEC and RM measurements. Two methods are tried and found to be largely successful. Both methods involve adding the missing density at heights close to the F2 peak. These are then essentially thin-shell models, and we also find that the standard thin-shell model, in which the entire TEC is placed in a thin layer at 350 km altitude, can reproduce the ionospheric contribution to the RM at a similar level, of order 0.06 rad m$^{-2}$. The advantage of the standard thin-shell model is that only GPS TEC measurements are required, and is computationally less expensive for determining the RM.

Additionally, we show that these measurements are best made locally at the observatory, and that standard global TEC models are much less successful at reproducing the measured data. We also show that the gradients of both observed and predicted RMs change at a rate of approximately 1 rad/m$^2$/hr, confirming that observations should be performed at intervals of at maximum 1.8 minutes in order to achieve a better than 0.03 rad/m$^2$ error, which is available through the GPS on site measurements, but not available through the global models.

We find that the intrinsic RM of the Pulsar B0950+08 to be 1.72 +/- 0.075 rad/m$^2$ based on the 3 days of observation from this study. This is in tension with the previous value of -0.66 +/- 0.04 rad/m$^2$ from \citet{Johnston}. We believe the authors failed to properly account for ionospheric variations in their study, which we have shown to be quite significant. 

Our conclusions are significant for measurements of the magnetic field in the solar wind and in coronal mass ejections using FR, where it is essential to remove the time-varying ionospheric contribution accurately. 

With pulsar measurement accuracy of 0.03 rad/m$^2$, the primary source of error in pulsar measurements remains in the ionosphere. We continue to push for new developments and new methods for describing the ionosphere to bring these levels below that of the pulsar measurement.

    \bibliographystyle{apalike}
    \bibliography{references}

\end{document}